\documentclass[10pt]{iopart}
\usepackage{iopams}

\usepackage{graphicx}
\usepackage[colorlinks=true]{hyperref}
\begin{document}

\title[Class.\ Quantum Grav.\ \textbf{36} (2019) 085007 (19pp)\qquad Y Furuya \textit{et al}]{A quantisation procedure in the presence of an initial Kasner singularity: primordial gravitational waves from triaxially anisotropic pre-inflation}

\author{Yu Furuya, Yuki Niiyama and Yuuiti Sendouda}
\address{
Graduate School of Science and Technology, Hirosaki University,\\
3 Bunkyocho, Hirosaki, Aomori 036-8561, Japan
}

\ead{
  \mailto{furuy(at)tap.st.hirosaki-u.ac.jp},
  \mailto{niiyama(at)tap.st.hirosaki-u.ac.jp},
  \mailto{sendouda(at)hirosaki-u.ac.jp}
}
\vspace{10pt}
\begin{indented}
\item[]4 April 2019
\end{indented}

\begin{abstract}
  In this paper, we discuss quantisation of cosmological tensor perturbations in the Kasner--de Sitter space-time as a model of (pre-)inflation.
  Quantisation in such an anisotropic background has been argued to be problematic based on the fact that the initial Kasner singularity, where the spatial anisotropy blows up, causes divergences to the effective frequencies squared for the perturbations, which render the standard quantisation procedure relying on the existence of an adiabatic vacuum state inexecutable.
  Here, an essential aspect of the problem is that the ability in determining the quantum spectra of the fields is restricted.
  Without its knowledge, one cannot even choose physically favourable states like the Bunch--Davies vacuum in de Sitter.
  We here argue that this difficulty may be circumvented if only there is a period, even if temporal, after the singularity where certain adiabatic conditions for the fields are met and the standard procedure of second quantisation can be carried out within the framework of the WKB approximation.
  We demonstrate that our prescription for determining the quantum energy spectrum is useful in making physically meaningful predictions for the primordial gravitational waves in triaxially anisotropic Kasner--de Sitter backgrounds.
  We confirm that, on short wave-length scales, the resulting spectrum and directional distribution of the primordial gravitational waves are the same as de Sitter inflation, namely, scale invariant and isotropic.
\end{abstract}

%
\vspace{2pc}
\noindent Keywords: cosmological perturbation theory, inflation, primordial gravitational waves\\
%
\submitto{\CQG}
%
%
%

\section{Introduction}

The original motivation for the proposal of cosmic inflation \cite{Guth:1980zm,Sato:1980yn}, an accelerated cosmological expansion in the very early universe, was to provide a solution to the horizon, flatness and magnetic-monopole problems.
Just after, it was realised that the quantum nature of inflation could even work as the seed of the various structures present in the late-time universe \cite{Hawking:1982cz,Starobinsky:1982ee,Guth:1982ec,Bardeen:1983qw}.
The inflationary theory has established its status as a paradigm of modern cosmology.

One of the unique predictions of the inflationary theories is generation of primordial gravitational waves (PGWs) during accelerated phases \cite{Mukhanov:1981xt}.
Detection of PGWs, direct or indirect, is considered to be a decisive evidence of the occurrence of primordial inflation.
Indirect search for PGWs is indeed one of the purposes of measuring the B-mode polarisations in the cosmic microwave background (CMB) \cite{Seljak:1996gy,Kamionkowski:1996zd,Kamionkowski:1996ks}, but its detection has not yet been accomplished.
Meanwhile, there has been a growing hope for directly detecting the PGWs since the great success of observing gravitational waves from binary celestial bodies by the LIGO and Virgo collaborations \cite{Abbott:2016blz,Abbott:2016nmj,Abbott:2017vtc,Abbott:2017gyy,Abbott:2017oio,TheLIGOScientific:2017qsa}.
Actually, direct detection of such PGWs in the low-frequency bands is one of the ultimate goals of future-planned laser-interferometer experiments in space such as LISA \cite{AmaroSeoane:2012je}, DECIGO/B-DECIGO \cite{Seto:2001qf,Nakamura:2016hna} and Big-Bang Observer \cite{Crowder:2005nr}.

Now, unveiling the origin and mechanism of cosmic inflation by means of observations should be the next target beyond confirmation of inflation.
When it comes to its origin, a primary issue should be whether or not there were preceding stages before onset of inflation;
If there was, it is well anticipated that the space-time before inflation was fairly anisotropic and/or inhomogeneous with a magnitude of the order of the energy scale of more fundamental theories such as grand unification theories (GUTs) or superstrings.
Searches for the remnants of those `pre-inflationary' anisotropies and inhomogeneities are well motivated.

Such pre-inflationary remnants, if any, cannot be probed easily with the currently available observations of CMB \cite{Ade:2015lrj,Akrami:2018odb}, which basically impose that the universe was isotropic at certain e-folds from the end of inflation.
Indeed, inflation is expected to have washed those anisotropies and inhomogeneities away;
anistropies of Bianchi-type universes \cite{Ellis:1968vb} are shown to be diminished by a cosmological constant \cite{Wald:1983ky} (but see \cite{Maleknejad:2012as} for what could happen in more realistic inflationary setup).

In this paper, we study a more direct probe---PGWs originating from a pre-inflationary stage.
The goal of our study will be to relate observables associated with PGWs to the primordial anisotropy before inflation, but, as an earlier attempt towards full generality, we here focus on anisotropies of Bianchi type-I, which is the simplest among the Bianchi types.

In a previous paper \cite{Furuya:2016dkh}, we performed a detailed analysis on classical aspects of PGWs in the Kasner--de Sitter (KdS) space-time, which has anisotropy of Bianchi type-I and is isotropised by a positive cosmological constant.
It was found that PGWs already existing in the pre-inflationary period can be amplified due to the effect of spatial shear before the isotropic de Sitter phase.

Now, we turn our attention to the quantum nature of PGWs originating from the anisotropic stage, since a full, consistent understanding, from its quantum origin to classical propagation, is necessary in order to derive reliable predictions.

Several observational consequences of anisotropic (pre-)inflation models in similar contexts were also studied in \cite{Gumrukcuoglu:2007bx,Pitrou:2008gk,Gumrukcuoglu:2008gi,Kofman:2011tr,Blanco-Pillado:2015dfa,Dey:2011mj,Dey:2012qp,Dey:2013tfa}, but many of them only focused on a particular type of anisotropy with so-called `oblate' axisymmetry.
This reflects the fact that there is an initial singularity in the presence of general, `triaxial' anisotropy, where Weyl tensor squared diverges.
The presence of singularity lies as a major obstacle when attempting to obtain reliable predictions from any calculations because identification of the unique quantum vacuum state is rendered difficult as discussed in \cite{Gumrukcuoglu:2007bx,Pitrou:2008gk}.

In this sense, understanding of the quantum nature of anisotropic universe should be considered still at an immature stage.
In particular, we have to make an effort to establish foundations of quantising fields in the presence of a general triaxial anisotropy.

The very purpose of this paper is to clarify in what manner and under which conditions reliable calculations associated with quantisation could be made.
Concretely, what we seek for is a prescription for determining the energy eigenstates.
We shall first revisit the problem of quantisation in anisotropic universe, illuminate the origin of the problem and propose a simple way to evade it.

This paper is organised as follows.
In section~\ref{sec:basic}, we show basic equations for the background and gravitational perturbations in the Kasner--de Sitter universe.
In section~\ref{sec:inicon}, we revisit the problem of quantisation in cosmological models with anisotropy of Bianchi type-I, and we present a prescription that helps to calculate quantum spectrum of tensor fluctuations under some conditions.
We also inspect applicability and limitation of our quantisation procedure.
In section~\ref{sec:PGWs}, we show the predicted power spectra and angular distributions of PGWs.
In section~\ref{sec:conc}, we conclude.

Throughout the paper, we use the natural units with $ c = \hbar = k_\mathrm B = 8 \pi G = 1 $\,.
Latin indices $ i, j, \cdots $ of vectors and tensors run through $ 1, 2, 3 $\,.

\section{\label{sec:basic}Basic equations}

In this section, we present basic equations for the background and gravitational perturbations in Kasner--de Sitter (KdS) universe in general relativity.
The formalism we employ was originally developed by Pereira \textit{et al.} \cite{Pereira:2007yy} for general Bianchi type-I cosmology (see \cite{Tomita:1985me} for an earlier attempt).
We adopt some notations used in a paper published by the authors \cite{Furuya:2016dkh}.
Readers who are familiar with this setup can skip this section.

\subsection{Background: Kasner--de Sitter solution}

Anisotropic cosmological models classified as Bianchi type-I \cite{Ellis:1968vb} have metric of the form
\begin{equation}
  g_{\mu\nu}\,\mathrm dx^\mu\,\mathrm dx^\nu
  =
  a(\eta)^2\,\left[
    -\mathrm d\eta^2
    + \sum_{i=1}^3 \left(\mathrm e^{\beta_i(\eta)}\,\mathrm dx^i\right)^2
  \right]\,,
  \label{eq:Bianchi}
\end{equation}
where $ a $ represents the \emph{average} scale factor and $ \beta_i $ ($ i = 1,2,3 $) the deviations from the average, satisfying $ \sum_{i=1}^3 \beta_i(\eta) = 0 $\,.
The spatial metric is defined as
\begin{equation}
  \gamma_{ij}
  = \mathrm{diag}(\mathrm e^{2\beta_1}\,,\mathrm e^{2\beta_2}\,,\mathrm e^{2\beta_3})\,,
  \quad
  \gamma^{ij}
  = \mathrm{diag}(\mathrm e^{-2\beta_1}\,,\mathrm e^{-2\beta_2}\,,\mathrm e^{-2\beta_3})\,,
\end{equation}
and used to raise and lower the indices of spatial tensors.

The Kasner--de Sitter space-time is a solution to Einstein's equations classified into this class, admitting positive cosmological constant $ \Lambda $ as the source.
The metric functions for the KdS solution are given by
\begin{equation}
  a(\eta)
  = a_\mathrm{iso}\,\sinh^{1/3}\left(3H_\Lambda t\right)\,,
  \quad
  \mathrm e^{\beta_i(\eta)}
  = \tanh^{q_i}\left(\frac{3H_\Lambda t}{2}\right)\,,
  \label{eq:aandbeta}
\end{equation}
where $ a_\mathrm{iso} $ is an arbitrary positive constant, $ H_\Lambda \equiv \sqrt{\Lambda/3} $\,, and $ t $ is the cosmic time related to $ \eta $ by $ \mathrm dt = a(\eta)\,\mathrm d\eta $\,.
The three exponents $ q_i $ ($ i = 1,2,3 $) are required to satisfy the constraints
\begin{equation}
  \sum_{i=1}^3 q_i
  = 0\,,
  \quad
  \sum_{i=1}^3 q_i^2
  = \frac{2}{3}\,.
  \label{eq:constraints}
\end{equation}
An angular parameter $ \Theta $ is introduced to express the exponents as
\begin{equation}
  q_1
  = \frac{2}{3}\,\sin\left(\Theta-\frac{2\pi}{3}\right)\,,
  \quad
  q_2
  = \frac{2}{3}\,\sin\left(\Theta-\frac{4\pi}{3}\right)\,,
  \quad
  q_3
  = \frac{2}{3}\,\sin\Theta\,.
\end{equation}

The magnitude of $ q_i $ quantifies the expansion rate along the $ x^i $-axis relative to the average.
Throughout the paper, we assume a hierarchy between the exponents
\begin{equation}
  q_1 \geq q_2 \geq q_3\,.
  \label{eq:hierarchy}
\end{equation}
This is done, without loss of generality, by restricting the range of $ \Theta $ to be $ 7\pi/6 \leq \Theta \leq 9\pi/6 $\,.
Due to the constraint (\ref{eq:constraints}) and the assumed hierarchy (\ref{eq:hierarchy}), the signs of $ q_1 $ and $ q_3 $ are fixed as $ q_1 > 0 > q_3 $\,.
In the triaxial case, $ q_1 > q_2 > q_3 $\,, the direction of the fastest expansion is associated to the $ x^1 $-axis and the slowest to the $ x^3 $-axis.
Two axisymmetric cases, $ (q_1,q_2,q_3) = (2/3,-1/3,-1/3) $ (`oblate') and $ (q_1,q_2,q_3) = (1/3,1/3,-2/3) $ (`prolate'), are realised for $ \Theta = 7\pi/6 $ and $ 9\pi/6 $\,, respectively.
In most of discussions in this paper, we take $ (q_1,q_2,q_3) = (1/\sqrt 3,0,-1/\sqrt 3) $ ($ \Theta = 8\pi/6 $) as most typical example with triaxial anisotropy.
The KdS solution has an initial singularity at $ t = 0 $ where a Weyl-curvature invariant $ C_{\mu\nu\rho\sigma}\,C^{\mu\nu\rho\sigma} $ diverges except for the oblate axisymmetric case.

The cosmic expansion is characterised by the average Hubble rate $ \mathcal H $ and the shear tensor $ \sigma_{ij} $ defined respectively by
\begin{eqnarray}
  \mathcal H
  &
    \equiv
    \frac{a'}{a}
    = a_\mathrm{iso}\,H_\Lambda\,\frac{\cosh(3H_\Lambda t)}{\sinh^{2/3}(3H_\Lambda t)}\,, \\
  \sigma_{ij}
  &
    \equiv
    \frac{1}{2}\,\gamma_{ij}'
    = 3 q_i\,a_\mathrm{iso}\,H_\Lambda\,
    \frac{\tanh^{2q_i}\left(\frac{3H_\Lambda t}{2}\right)}
    {\sinh^{2/3}(3H_\Lambda t)}\,
    \delta_{ij}\,,
\end{eqnarray}
where the prime denotes differentiation with respect to $ \eta $\,.
The KdS universe has two regimes distinguished by the magnitude of the shear.
In the earlier, `Kasner-like' period for $ t \ll H_\Lambda^{-1} $\,, the average scale factor $ a $ and the spatial metric $ \gamma_{ij} $ are approximated as
\begin{equation}
  a(\eta) 
  \sim
  a_\mathrm{iso}\,(3H_\Lambda t)^{1/3}\,,
  \quad 
  \gamma_{ij}(\eta)
  \sim
  \left(\frac{3H_\Lambda t}{2}\right)^{2q_i}\,\delta_{ij}\,.
\end{equation}
On the other hand, in the later, `de Sitter-like' period for $ t \gg H_\Lambda^{-1} $\,, $ a $ and $ \gamma_{ij} $ are approximated as
\begin{equation}
  a(\eta)
  \sim
  2^{-1/3}\,a_\mathrm{iso}\,\mathrm e^{H_\Lambda t}\,,
  \quad
  \gamma_{ij}(\eta)
  \sim
  \delta_{ij}\,.
\end{equation}
The time $ t = t_\mathrm{iso} \equiv H_\Lambda^{-1} $ appoints a moment around which the transition from the anisotropic to the isotropic regime occurs.

\subsection{Polarisation basis}

To perform scalar-vector-tensor decomposition of metric perturbations, we need to introduce a set of polarisation basis associated with the wave vector $ \vec k $ of perturbations.
As usual, we will parameterise the harmonic modes of waves by a set of constants $ (k_1,k_2,k_3) $, which are regarded as the covariant components of a wave vector in the $ (x^1,x^2,x^3) $ comoving coordinate frame.
In an anisotropic background, the contravariant components defined as $ k^i \equiv k_j\,\gamma^{ij} $ are not always constant, and wave vector $ \vec k \equiv (k^1,k^2,k^3) $ changes its direction and norm in time during the anisotropic regime, $ 0 < t \lesssim t_\mathrm{iso} $ \cite{Pereira:2007yy}.
Only after isotropisation of the universe the wave vector coincides with its covariant dual, i.e.\ $ \lim_{t\to\infty} \vec k = (k_1,k_2,k_3) $\,.

Then we introduce the polarisation-basis vectors $ (\vec{\hat k}\,, \vec e_{(1)}\,, \vec e_{(2)}) $\,, where $ \vec{\hat k} \equiv \vec k/\sqrt{\gamma^{ij}\,k_i\,k_j} $ is the normalised wave vector, and $ \vec e_{(1)}\,, \vec e_{(2)} $ are the orthonormal polarisation vector basis perpendicular to $ \vec{\hat k} $\,.
Like $ \vec k $\,, the basis vectors change their direction in the anisotropic regime.
They are usefully viewed as rigidly rotated orthonormal basis in the coordinate frame in terms of time-dependent Euler angles $ (\alpha,\beta,\gamma) $ as defined in \cite{Pitrou:2008gk}.
The angles $ (\beta,\gamma) $ correspond to the zenith angle and azimuth angle of $ \vec{\hat k} $\,, respectively.
Actually, the normalised wave vector $ \vec{\hat k} $ is parameterised as
\begin{equation}
  \vec{\hat k}
  =
  \left(
    \begin{array}{c}
      \mathrm e^{-\beta_1}\,\sin\beta\,\cos\gamma \\
      \mathrm e^{-\beta_2}\,\sin\beta\,\sin\gamma \\
      \mathrm e^{-\beta_3}\,\cos\beta \\
    \end{array}
  \right)\,.
  \label{eq:k}
\end{equation}
Also, the other basis vectors can be expressed as
\begin{eqnarray}
  \vec e_{(1)}
  &
    \equiv
    \left(
    \begin{array}{c}
      \mathrm e^{-\beta_1}\,
      (\cos\beta\,\cos\gamma\,\cos\alpha - \sin\gamma\,\sin\alpha) \\
      \mathrm e^{-\beta_2}\,
      (\cos\beta\,\sin\gamma\,\cos\alpha + \cos\gamma\,\sin\alpha) \\
      -\mathrm e^{-\beta_3}\,\sin\beta\,\cos\alpha
    \end{array}
  \right)\,, \\
  \vec e_{(2)}
  &
    \equiv
    \left(
    \begin{array}{c}
      -\mathrm e^{-\beta_1}\,
      (\cos\beta\,\cos\gamma\,\sin\alpha + \sin\gamma\,\cos\alpha) \\
      -\mathrm e^{-\beta_2}\,
      (\cos\beta\,\sin\gamma\,\sin\alpha - \cos\gamma\,\cos\alpha) \\
      \mathrm e^{-\beta_3}\,
      \sin\beta\,\sin\alpha
    \end{array}
  \right)\,,
\end{eqnarray}
where $ \alpha $ is another time-dependent angle corresponding to the residual rotation degree of freedom of $ (\vec e_{(1)},\vec e_{(2)}) $ around $ \vec{\hat k} $\,.
As in reference~\cite{Pitrou:2008gk}, we require $ \alpha $ satisfies a constraint condition
\begin{equation}
  \alpha'
  =
  -\gamma'\,\cos\beta\,.
\end{equation}
The tensor basis are defined in terms of the vector basis as
\begin{equation}
  \varepsilon^+_{ij} 
  \equiv
  \frac{e^{(1)}_i\,e^{(1)}_j - e^{(2)}_i\,e^{(2)}_j}{\sqrt 2}\,,
  \quad
  \varepsilon^\times_{ij}
  \equiv
  \frac{e^{(1)}_i\,e^{(2)}_j + e^{(2)}_i\,e^{(1)}_j}{\sqrt 2}\,.
\end{equation}
where indices $ +, \times $ indicates the two degrees of freedom of gravitational waves.

Any scalar and tensor functions are decomposed into Fourier modes as
\begin{eqnarray}
  f(x_i,\eta)
  &
    =
    \int\frac{\mathrm d^3k_i}{(2\pi)^{3/2}}
    \tilde{f}(k_i,\eta)\,\mathrm e^{\mathrm i\,k_j\,x^j}\,, \\
  V_{ij}(x_i,\eta)
  &
    =
    \sum_{\lambda=+,\times} \int\frac{\mathrm d^3k_i}{(2\pi)^{3/2}}
    \tilde{V}_\lambda(k_i,\eta)\,\mathrm e^{\mathrm i\,k_l\,x^l}\,\varepsilon^\lambda_{ij}(k_i,\eta)\,.
\end{eqnarray}
For instance, the shear tensor $ \sigma_{ij} $ is decomposed into the scalar $ \sigma^{(\mathrm S)} \equiv \sigma_{ij}\,\hat k^i\,\hat k^j $\,, vector $ \sigma^{(\mathrm V)}_{(a)} \equiv \sigma_{ij}\,\hat k^i\,e_{(a)}^j$ ($ a = 1,2 $), and tensor part $ \sigma^{(\mathrm T)}_\lambda \equiv \sigma_{ij}\,\varepsilon_\lambda^{ij} $ ($ \lambda = +,\times $)\,, respectively, by the polarisation basis introduced above.

\subsection{Equations of motion and action for gravitational waves}

In the gauge-invariant formalism by Pereira \textit{et al.} \cite{Pereira:2007yy}, the general perturbed metric is given by
\begin{eqnarray}
  &
    (g_{\mu\nu} + \delta g_{\mu\nu})\,\mathrm dx^{\mu}\,\mathrm dx^{\nu} \nonumber\\
  & \quad
    =
    a(\eta)^2\,
    \left[
    -(1 + 2A)\,\mathrm d\eta^2
    + 2 B_i\,\mathrm dx^i\,\mathrm d\eta
    + (\gamma_{ij} + h_{ij})\,\mathrm dx^i\,\mathrm dx^j
    \right]\,,
\end{eqnarray}
where the $ (0i) $ and $ (ij) $ components are respectively decomposed into the scalar, vector, and tensor variables as
\begin{eqnarray}
  B_i
  &
    =
    \partial_i B + \bar B_i\,, \\
  h_{ij}
  &
    =
    2\,(\gamma_{ij} + \mathcal H^{-1}\,\sigma_{ij})\,C
    + 2 \partial_{ij}E + 2 \partial_{(i}E_{j)} + 2 E_{ij}\,,
\end{eqnarray}
where the vector and tensor variables satisfy $ \partial_i \bar B^i = \partial_i E^i = \partial_i E^i{}_j = E^i{}_i = 0 $\,.
In vacuum (plus a cosmological constant), the only dynamical degrees of freedom are represented by the two polarisation components of the gauge-invariant tensor variable $ E_{ij} $ defined by
\begin{equation}
  E_\lambda(k_i,\eta)
  \equiv
  \int\!\frac{\mathrm d^3x}{(2\pi)^{3/2}}\,E_{ij}(x_i,\eta)\,\mathrm e^{-\mathrm i\,k_l\,x^l}\,
  \varepsilon_\lambda^{ij}
  \quad
  (\lambda = +,\times)\,.
\end{equation}
Introducing $ \mu_\lambda \equiv a\,E_\lambda $\,, the equations of motion for gravitational waves are given as \cite{Pereira:2007yy}
\begin{eqnarray}
  &
    \mu_+'' + \omega_+^2\,\mu_+ + \xi\,\mu_\times = 0\,, \label{eq:mupeom} \\
  &
    \mu_\times'' +\omega_\times^2\,\mu_\times + \xi\,\mu_+ = 0\,, \label{eq:muceom}
\end{eqnarray}
where
\begin{eqnarray}
  \omega_+^2
  &
    \equiv
    \gamma^{ij}\,k_i\,k_j
    - \frac{a''}{a} \nonumber\\
  & 
    \quad
    - \frac{\left(a^2\,\sigma^{(\mathrm S)}\right)'}{a^2}
    - 2\,\left(\sigma_\times^{(\mathrm T)}\right){}^2
    - \frac{2}{a^2}\,\left(
    \frac{a^2\,\left(\sigma_+^{(\mathrm T)}\right){}^2}
    {2 \mathcal H-\sigma^{(\mathrm S)}} 
    \right)'\,, \\
  \omega_\times^2
  &
    \equiv
    \gamma^{ij}\,k_i\,k_j
    - \frac{a''}{a} \nonumber\\
  &
    \quad
    - \frac{\left(a^2\,\sigma^{(\mathrm S)}\right)'}{a^2}
    - 2\,\left(\sigma_+^{(\mathrm T)}\right){}^2
    - \frac{2}{a^2}\,\left(
    \frac{a^2\,\left(\sigma_\times^{(\mathrm T)}\right){}^2}
    {2 \mathcal H-\sigma^{(\mathrm S)}} 
    \right)'\,,
\end{eqnarray}
and
\begin{equation}
  \xi
  \equiv 
  2 \sigma_+^{(\mathrm T)}\,\sigma_\times^{(\mathrm T)}
  - \frac{2}{a^2}\,\left( 
    \frac{a^2\,\sigma_+^{(\mathrm T)}\,\sigma_\times^{(\mathrm T)}}
    {2 \mathcal H-\sigma^{(\mathrm S)}} 
  \right)'\,.
\end{equation}
Correspondingly, the second-order action for $ \mu_\lambda $ was also obtained by Pereira \textit{et al.} \cite{Pereira:2007yy}.
Discarding scalar fields in equation~(5.20) in \cite{Pereira:2007yy}, one obtains the action integral for tensor perturbations
\begin{equation}
  S_2
  =
  \sum_{\lambda=+,\times}
  \frac{1}{2}\,\int\!\mathrm d^3k_i\,\mathrm d\eta\,
  \left[
    \mu'_\lambda\,\mu'_\lambda{}^*
    - \omega_\lambda^2\,\mu_\lambda\,\mu_\lambda{}^*
    - \xi\,\mu_\lambda^*\,\mu_{1-\lambda}
  \right]\,,
  \label{eq:muaction}
\end{equation}
where the index `$ 1-\lambda $' is $ \times $ for $ \lambda = + $ and $ + $ for $ \lambda = \times $\,.

\section{\label{sec:inicon}Quantisation of PGWs in triaxial Kasner--de Sitter universe}

In this section, we consider quantisation of tensor perturbations in the triaxially anisotropic KdS background.

Since the action for the tensor perturbations (\ref{eq:muaction}) has an analogous shape to (coupled) double scalar fields, one might think that an ordinary procedure of quantisation should work.
This is simply not the case, though, for $ \omega_\lambda^2 $ generically blows up due to the growing background anisotropy as approaching to the initial singularity.

To see this, it is instructive to recall how the contravariant wave vector $ \vec k $ rotates under the influence of anisotropic expansion.
As going back in time towards the initial singularity, any wave vector $ \vec k $ with non-vanishing $ k^1 $ component tends to be aligned to the $ k^1 $-axis, as one can see the limits of the Euler angles using (\ref{eq:k})\,:
\begin{eqnarray}
  \beta
  &
  = \cos^{-1}\left(\frac{k_3\,\mathrm e^{-\beta_3}}{\sqrt{\sum_{i=1}^3 k_i^2\,\mathrm e^{-2\beta_i}}}\right)
  \to \frac{\pi}{2}\,, \\
  \gamma
  &
  = \tan^{-1}\left(\frac{k_2\,\mathrm e^{-\beta_2}}{k_1\,\mathrm e^{-\beta_1}}\right)
  \to 0\,.
\end{eqnarray}
In such an asymptotic regime, the frequency squared $ \omega_\lambda^2 $ can be estimated approximating $ k_2 \sim k_3 \sim 0 $ as
\begin{equation}
  \omega_\lambda^2
  \sim
  k_1^2\,\left(\frac{3 H_\Lambda t}{2}\right)^{-2q_1}
  + a_\mathrm{iso}^2\,H_\Lambda^2\,
  \frac{
    1-9\,\delta^\times_\lambda\,\left(q_2-q_3\right)^2
  }{(3 H_\Lambda t)^{4/3}}\,.
  \label{eq:omegasq}
\end{equation}
This expression illuminates the generic divergent behaviour of $ \omega_\lambda^2 $ as $ t \to 0 $\,.

The above observation implies that the variable $ \mu_\lambda $ is not the best quantity to quantise, but we should seek for a more suitable variable which can behave like a simple harmonic oscillator.

Now, we revisit the problems of quantisation of primordial perturbations in the Kasner space-time.
They were discussed in reference~\cite{Pitrou:2008gk}, and here, we augment them in terms of a little more accurate expressions.
Then we introduce a prescription for quantisation which may enable us to evade the problem.

\subsection{Problems in quantisation}

The usual procedure of second quantisation in cosmological background relies upon the existence of suitable modes which oscillate harmonically in the asymptotic past region.
In an anisotropic setup, however, the spatial shear generally dominates near the singularity at $ t = 0 $ and often causes instabilities of negative frequency in each mode \cite{Gumrukcuoglu:2007bx,Pereira:2007yy,Pitrou:2008gk,Kofman:2011tr}, as inferred by (\ref{eq:omegasq}).

In order to gain some insights, it is useful to rewrite the equations of motion for $ \mu_\lambda $ (\ref{eq:mupeom}) (\ref{eq:muceom}) in terms of the new variable and time coordinate \cite{Pitrou:2008gk}
\begin{equation}
  \chi_\lambda = f(\eta)\,\mu_\lambda\,,
  \quad
  \mathrm d\tau = f(\eta)^2\,\mathrm d\eta\,.
\end{equation}
With the definitions
\begin{equation}
  \Omega_\lambda^2
  \equiv
  \frac{\omega_\lambda^2}{f^4}
  + \frac{(f^{-1})''}{f^3}\,,
  \quad
  \Xi
  \equiv
  \frac{\xi}{f^4}\,,
\end{equation}
the equations of motion for $ \chi_\lambda $ read
\begin{eqnarray}
  &
    \ddot\chi_+
    + \Omega_+^2\,\chi_+
    + \Xi\,\chi_\times
    = 0\,, \\
  &
    \ddot\chi_\times
    + \Omega_\times^2\,\chi_\times
    + \Xi\,\chi_+
    = 0\,,
\end{eqnarray}
where the dots denote derivatives with respect to $ \tau $\,.
Here, apart from the subtlety of the interaction term, there arises a possibility of making $ \Omega_\lambda^2 $ constant by choosing some function $ f $ so that $ \chi_\lambda $ behaves as a harmonic oscillator.

Let us consider a generic power-law function $ f(\eta) = \left(a_\mathrm{iso} H_\Lambda \eta\right)^p $ as in \cite{Pitrou:2008gk}.
Then the new frequency squared $ \Omega_\lambda^2 $ is evaluated near the singularity, unless $ k_1 = 0 $\,, as
\begin{eqnarray}
  \Omega_\lambda^2
    \sim
    \frac{k_1^2}{2^{q_1}\,(a_\mathrm{iso} H_\Lambda \eta)^{4p+3q_1}}
    + a_\mathrm{iso}^2\,H_\Lambda^2\,
    \frac{(2p+1)^2 - 9\,\delta^\times_\lambda\,(q_2-q_3)^2}{4\,(a_\mathrm{iso} H_\Lambda \eta)^{4p+2}}\,.
    \label{eq:Omegaetap}
\end{eqnarray}

The oblate axisymmetric KdS, $ (q_1,q_2,q_3) = (2/3,-1/3,-1/3) $\,, is the exceptional case where the powers of the terms in $ \Omega_\lambda^2 $ becomes identical.
Moreover, by choosing $ p = -1/2 $\,, it becomes \emph{positive} constant $ \Omega_\lambda^2 \sim k_1^2 $ unless $ k_1 = 0 $\,.
Namely, in this case, quantisation can be carried out except for the `planar modes' with $ k_1 = 0 $\,.
For this reason, several authors have studied primordial perturbations of quantum origin in this particular symmetric background \cite{Gumrukcuoglu:2007bx,Kim:2010wra,Dey:2011mj,Dey:2012qp,Dey:2013tfa,Blanco-Pillado:2015dfa}.

In any other cases of background, all the indices $ q_i $ ($ i = 1,2,3 $) are smaller than $ 2/3 $\,, so the second term in (\ref{eq:Omegaetap}) always dominates.
Then, setting $ p = -1/2 $ eliminates the time dependence in the leading term, but this leads to a \emph{negative} frequency squared for the cross mode ($ \lambda = \times $)
\begin{equation}
  \Omega_\times^2
  \sim  
  -\frac{9\,(q_2-q_3)^2}{4}\,(a_\mathrm{iso}\,H_\Lambda)^2\,,
\end{equation} 
which forbids the usual quantisation procedure.

The above analysis supplements the considerations in \cite{Pitrou:2008gk}, in which they focused upon adiabaticity conditions.

Now, one realises that the problem just arises from the ordinary prescription of quantisation that is founded on asymptotic behaviours of the mode functions, which is ill-behaved in the current setup due to the existence of the initial singularity.
However, the initial singularity is often a source of difficulties and its existence is disfavoured on the physical background.
It is rather natural to suppose that the history of the universe began with some finite curvature at some moment $ t = t_\mathrm{ini} > 0 $\,, and quantisation of the field contents is done at or after $ t_\mathrm{ini} $\,.

If we introduce a finite initial time $ t_\mathrm{ini} $\,, then there arises a chance that the wave-number terms dominate in $ \omega_\lambda^2 $ at or after $ t_\mathrm{ini} $\,, and the frequency squared after transformation may be evaluated as
\begin{equation}
  \Omega_\lambda^2
  \sim
  \sum_{i=1}^3 \frac{k_i^2}{2^{q_i}\,(a_\mathrm{iso} H_\Lambda \eta)^{4p+3q_i}}\,.
\end{equation}
Once the dominant term among the three components above, to be indicated by the index $ i_\mathrm{max} $, is identified, setting $ p = -3\,q_{i_\mathrm{max}}/4 $ results in a positive, constant frequency squared $ \Omega_\lambda^2 \sim k_{i_\mathrm{max}}^2 $\,.
Note that this is analogous to the case of `oblate' axisymmetric backgrounds, where quantisation can be carried out except for the `planar' modes \cite{Gumrukcuoglu:2007bx,Kim:2010wra,Dey:2011mj,Dey:2012qp,Dey:2013tfa,Blanco-Pillado:2015dfa}.

In what follows, we proceed this idea and give a viable prescription of quantisation in triaxially anisotropic universe.

\subsection{\label{sec:trans}Transformations}

For the sake of brevity, we collectively call the following three functions of time (and not of wave number) as $ {}^{(i)}f $ ($ i = 1,2,3 $):
\begin{eqnarray}
  {}^{(1)}f
  &
    =
    \tanh^{-q_1/2}\left(\frac{3 H_\Lambda t}{2}\right)\,, \\
  {}^{(2)}f
  &
    =
    \tanh^{-q_2/2}\left(\frac{3 H_\Lambda t}{2}\right)\,, \\
  {}^{(3)}f
  &
    = \tanh^{-q_3/2}\left(\frac{3 H_\Lambda t}{2}\right)\,.
\end{eqnarray}
With them, the wave-number terms in $ \omega_\lambda^2 $ can be written as
\begin{equation}
  \gamma^{ij}\,k_i\,k_j = {}^{(1)}f^4\,k_1^2 + {}^{(2)}f^4\,k_2^2 + {}^{(3)}f^4\,k_3^2\,.
\end{equation}
Since the factors $ {}^{(i)}f^4 $ depend on time to the distinctive powers in a triaxially anisotropic background ($ q_1 > q_2 > q_3 $), only one of the above three terms should dominate at any moment.
Keeping this in mind, let us consider a transformation using $ {}^{(i)}f $ as the transformation function $ f $\,:
\begin{equation}
  {}^{(i)}\chi_\lambda
  = {}^{(i)}f\,\mu_\lambda\,,
  \quad
  \mathrm d{}^{(i)}\tau
  = {}^{(i)}f^2\,\mathrm d\eta\,.
  \label{eq:chiandtau}
\end{equation}
Then the frequency squared for the new variable $ {}^{(i)}\chi_\lambda $ is written as
\begin{eqnarray}
  {}^{(i)}\Omega_\lambda^2
  &
    =
    \frac{\omega_\lambda^2}{{}^{(i)}f^4} + \frac{({}^{(i)}f^{-1})''}{{}^{(i)}f^3} \\
  &
    =
    k_i^2
    + \sum_{j\neq i} k_j^2\,\tanh^{-2(q_j - q_i)}\left(\frac{3 H_\Lambda t}{2}\right)
    + \cdots\,,
\end{eqnarray}
where only the terms relevant to the norm of wave vector were shown in the second line.
Now we introduce the time for quantisation $ t = t_* $\,, which is not necessarily $ t_\mathrm{ini} $ but can be any moment after it.
Then, once a wave vector $ (k_1,k_2,k_3) $ is given, one can decide whether each of the functions $ \{{}^{(1)}f,{}^{(2)}f,{}^{(3)}f\} $ can satisfy the condition
\begin{equation}
  k_i^2
  \gtrsim
  \sum_{j\neq i} k_j^2\,\tanh^{-2(q_j - q_i)}\left(\frac{3 H_\Lambda t_*}{2}\right)\,,
\end{equation}
and it is under this condition that a variable $ {}^{(i)}\chi_\lambda $ has an approximately constant frequency squared $ {}^{(i)}\Omega_\lambda^2 \sim k_i^2 $\,.
We regard such a variable as `suitable for quantisation'.

\begin{figure}[htbp]
  \includegraphics[scale=0.70]{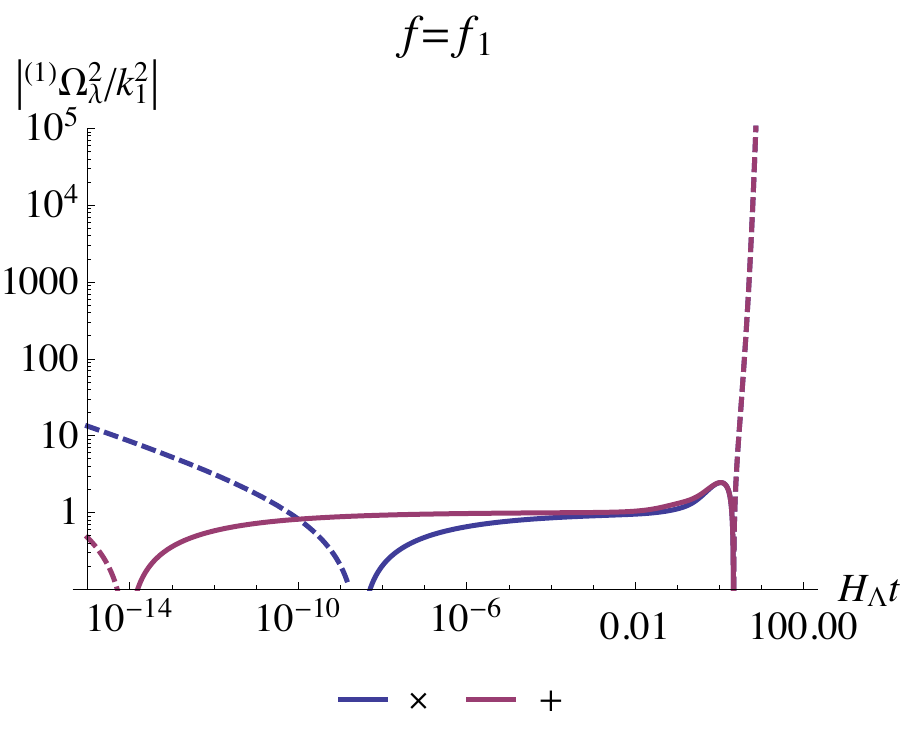}
  \includegraphics[scale=0.70]{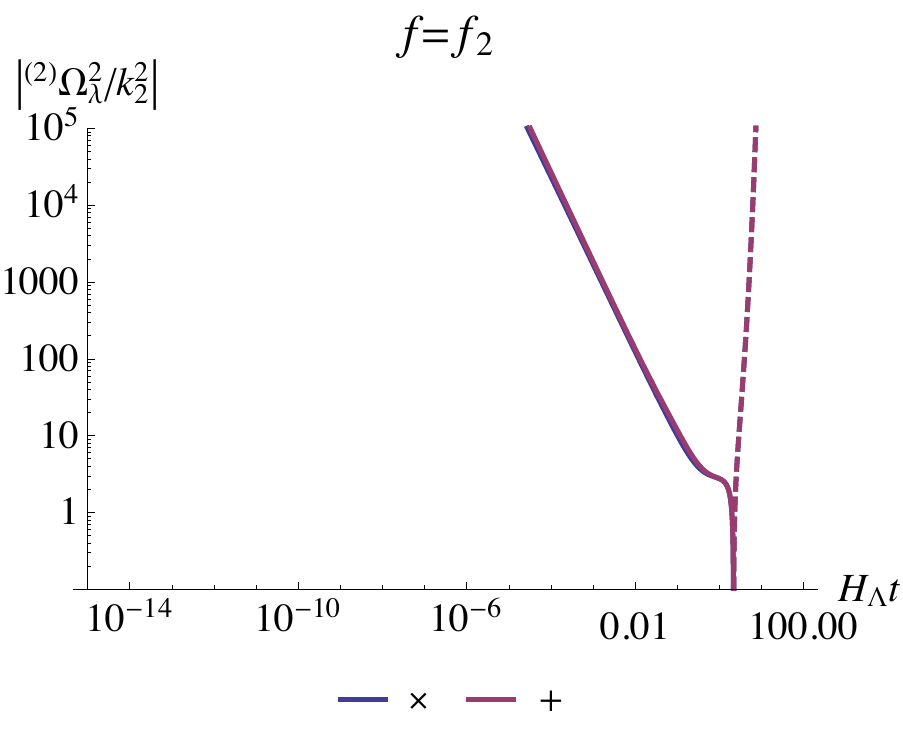}
  \includegraphics[scale=0.70]{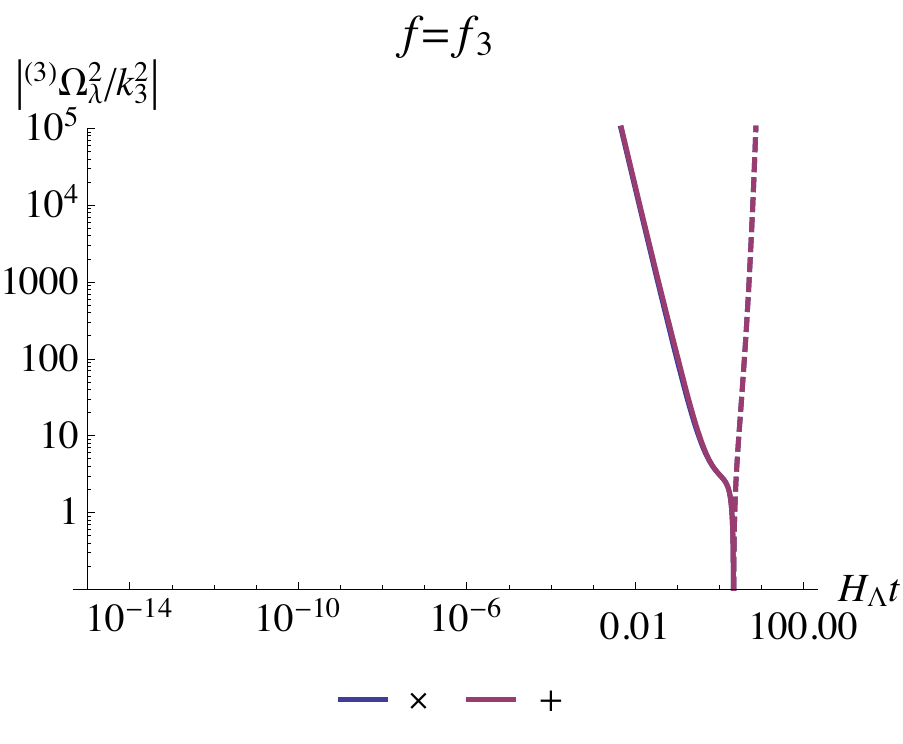}
  \caption{\label{fig:Omega}Time evolution of the frequencies squared for $ {}^{(i)}f $-transformed variables $ {}^{(i)}\chi_\lambda $\,: $ {}^{(1)}\Omega_\lambda^2 $ (top-left), $ {}^{(2)}\Omega_\lambda^2 $ (top-right) and $ {}^{(3)}\Omega_\lambda^2 $ (bottom).
    The mode has a wave number $ k_1 = k_2 = k_3 = (10/\sqrt 3)\,a_\mathrm{iso}\,H_\Lambda $ and the background is triaxially anisotropic with the exponents $ (q_1,q_2,q_3) = (1/\sqrt 3,0,-1/\sqrt 3) $ ($ \Theta = 8\pi/6 $).
    The lines are dashed where the values are negative.}
\end{figure}

Figure~\ref{fig:Omega} shows the time evolution of the squared frequencies $ {}^{(i)}\Omega_\lambda^2 $ for the transformed variables $ {}^{(i)}\chi_\lambda = {}^{(i)}f\,\mu_\lambda $ ($ i = 1,2,3 $) with a wave number $ k_1 = k_2 = k_3 = (10/\sqrt 3)\,a_\mathrm{iso}\,H_\Lambda $\,.
The background is triaxially anisotropic, with the exponents $ (q_1,q_2,q_3) = (1/\sqrt 3,0,-1/\sqrt 3) $ ($ \Theta = 8\pi/6 $).
In this case, $ {}^{(1)}\Omega_\lambda^2 $ stays constant in a wide range of time (top-left), whereas $ {}^{(2)}\Omega_\lambda^2 $ and $ {}^{(3)}\Omega_\lambda^2 $ are not (top-right and bottom).
This tendency is a consequence of the fact that the wave-number term in $ \omega_\lambda^2 $ can be approximated as
\begin{equation}
  \gamma^{ij}\,k_i\,k_j
  \simeq
  {}^{(1)}f^4\,k_1^2
  \quad
  (t \ll t_\mathrm{iso}\,, q_1 > q_2 > q_3)
\end{equation}
thanks to the hierarchy $ {}^{(1)}f \gg {}^{(2)}f \gg {}^{(3)}f $\,.
This applies to a wide range of the choice of wave vector $ (k_1,k_2,k_3) $\,.

The figures illuminate that the advantage of taking $ {}^{(1)}\chi_\lambda $ as a quantity to quantise, since it would oscillate harmonically for the longest period.
However, as seen in the figure, $ {}^{(1)}\Omega_\lambda^2 $ ceases to stay constant for smaller $ t $\,, where the terms other than the wave number become dominant.
As discussed in the previous section, this is inevitable if a mode really approaches to the initial singularity at $ t = 0 $\,.
Quantitatively, in this particular but typical example, we may conclude that $ {}^{(1)}\chi_\lambda $ should be the most sensible choice as a variable for quantisation as long as we quantise at some moment between $ t = 10^{-6}\,H_\Lambda^{-1} $ and $ 10^{-2}\,H_\Lambda^{-1} $\,. 

Likewise, once the time of quantisation is given, we examine whether or not each variable $ {}^{(i)}\chi_\lambda $ ($ i = 1,2,3 $) is suitable for quantisation.
We suggest the following working criterion for the usefulness of a variable $ {}^{(i)}\chi_\lambda $ in quantisation:
If
\begin{equation}
  {}^{(i)}\Omega_\lambda^2 > 0
\end{equation}
and
\begin{equation}
  {}^{(i)}R_\lambda
  \equiv
  \left|
    \frac{1}{{}^{(i)}\Omega_\lambda^2}\,
    \frac{\mathrm d{}^{(i)}\Omega_\lambda}{\mathrm d{}^{(i)}\tau}
  \right|
  \ll 1
\end{equation}
are satisfied simultaneously, $ {}^{(i)}\chi_\lambda $ is a sensible choice of variable for quantisation.

\begin{figure}[htbp]
  \begin{center}
  \includegraphics[scale=0.4]{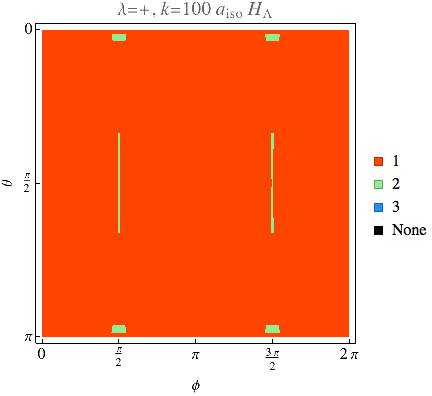}
  \includegraphics[scale=0.4]{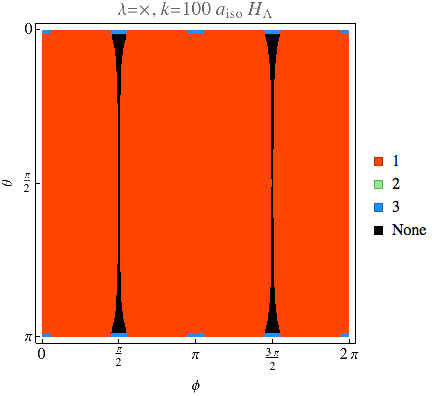}\\
  \includegraphics[scale=0.4]{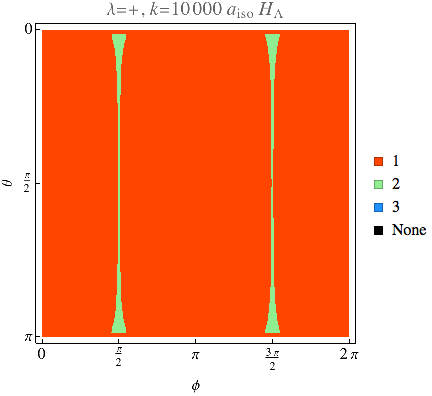}
  \includegraphics[scale=0.4]{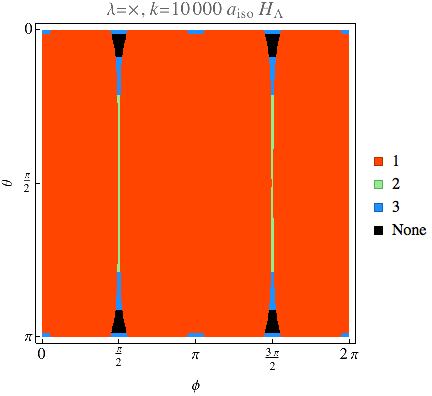}
  \end{center}
  \caption{\label{fig:R}Showing of which $ {}^{(i)}R_\lambda $ ($ i = 1,2,3 $) is the smallest for each mode pointing the direction specified by the angular coordinates $ (\theta,\phi) $ in the celestial sphere.
    Regions labelled by `$ i $' indicate $ {}^{(i)}R_\lambda $ is the smallest, whereas in the regions labelled by `None', $ {}^{(i)}\Omega^2_\lambda < 0 $ for $ \forall i $\,.
    The background anisotropy is triaxial ($ \Theta = 8\pi/6 $) and the wave number is $ k = 10^2\,a_\mathrm{iso}\,H_\Lambda\,, 10^4\,a_\mathrm{iso}\,H_\Lambda $ from top to bottom.}
\end{figure}

Figure~\ref{fig:R} shows which of $ \{{}^{(1)}R_\lambda,{}^{(2)}R_\lambda,{}^{(3)}R_\lambda\} $ is the smallest for each direction of $ (k_1,k_2,k_3) $ with several wave numbers $ k \equiv \sqrt{k_1^2 + k_2^2 + k_3^2} $\,.
These figures are the Mercator projection of the `celestial sphere' parameterised by the polar angle $ \theta $ and azimuth angle $ \phi $ defined as
\begin{eqnarray}
  k_1
  &
    = k\,\sin\theta\,\cos\phi\,, \\
  k_2
  &
    = k\,\sin\theta\,\sin\phi\,, \\
  k_3
  &
    = k\,\cos\theta\,.
\end{eqnarray}
In the figures, regions labelled with `$ i $' ($ i = 1,2,3 $) correspond to where $ {}^{(i)}R_\lambda $ is the smallest of the three.
For the cross mode ($ \lambda = \times $, right column), there appear regions labelled as `None' (black).
For those wave numbers, none of the three frequencies squared $ {}^{(i)}\Omega_\lambda^2 $ is positive, which implies none of the variables $ {}^{(i)}\chi_\lambda $ is suitable for quantisation.
This does not necessarily mean quantisation is impossible, for it merely alerts the failure of our restricted procedure, but in the current work, we would not proceed to quantise such modes.

We see that the $ {}^{(1)}f $-transformed variable $ {}^{(1)}\chi_\lambda $ is the most suitable choice for quantisation in a vast area on the celestial sphere.
This is a direct consequence of the hierarchy $ {}^{(1)}f \gg {}^{(2)}f \gg {}^{(3)}f $ for $ t \ll H_\Lambda^{-1} $\,.
Other variables could take over only when $ k_1 $ is tiny, in the vicinity of the circumferences with $ \phi = \pi/2, 3\pi/2 $ where $ k_1 = 0 $\,.
In the almost whole region on those circumferences, $ {}^{(2)}\chi_\lambda $ is superior to $ {}^{(3)}\chi_\lambda $ because of, again, the above hierarchy.

\subsection{\label{sec:conds}Conditions for sensible quantisation}

To sum up, we expect that quantisation in the triaxially Kasner--de Sitter universe can be carried out to good approximation if there is some moment $ t_* $ around which the following conditions are fulfilled:
\begin{itemize}
\item
  At least one of the transformed variables $ \{{}^{(1)}\chi_\lambda\,, {}^{(2)}\chi_\lambda\,, {}^{(3)}\chi_\lambda\} $ has positive, nearly constant frequency squared.
\item
  Interaction between the polarisation modes are negligible.
  Decoupling of the modes in the short wave-length regime has been partly confirmed in \cite{Pereira:2007yy}, and we expect that the interaction is sufficiently weak as long as the WKB approximation is valid.
\end{itemize}

The role of the above conditions is only to provide us the means to determine the energy spectrum of the fields at $ t = t_* $\,.
Here, to choose an adequate quantum state is a completely different problem which we would not answer at this stage.
A typical situation is that, although the state with the lowest energy at $ t = t_* $ is a tempting choice, it may not remain the ground state throughout the whole anisotropic expansion unlike the Bunch--Davies vacuum in isotropic de Sitter.
We can however expect that the knowledge of the energy spectrum should serve as a useful clue to choose the favoured state on physical backgrounds.

We shall see the result of quantisation of PGWs under these conditions in section~\ref{sec:PGWs}.
In the following discussions, we assume that we can decide which of $ {}^{(i)}\chi_\lambda $ is the most suitable choice of a variable to quantise for a given wave vector $ (k_1,k_2,k_3) $ within a given setup.
With this in mind, the index $ (i) $ is omitted unless it is necessary.

\subsection{Canonical quantisation}

Once a variable suitable for quantisation is given, a standard procedure of second quantisation can be carried out.
What we need is the normalisation condition imposed on each mode.

The second-order action for $ \chi_\lambda $ is given by transforming the second-order action for $ \mu_\lambda $ (\ref{eq:muaction}) as
\begin{eqnarray}
  S_2
  &
    \simeq
    \sum_{\lambda=+,\times}
    \frac{1}{2}\,\int\!\mathrm d^3k_i\,\mathrm d\eta\,
    \left[
    \mu'_\lambda\,\mu'_\lambda{}^*
    - \omega_\lambda^2\,\mu_\lambda\,\mu_\lambda{}^*
    \right] \\
  &
    =
    \sum_{\lambda=+,\times}
    \frac{1}{2}\,\int\!\mathrm d^3k_i\,\mathrm d\tau\,
    \left[
    \dot\chi_\lambda\,\dot\chi_\lambda{}^*
    - \Omega_\lambda^2\,\chi_\lambda\,\chi_\lambda{}^*
    \right]
\end{eqnarray}
up to surface terms, where we have dropped the interaction term.
The operator version of the canonical variable is
\begin{equation}
  \hat\chi_\lambda
  = \hat a^\lambda_{k_i}\,u_\lambda\,,
\end{equation}
where $ \hat a_\lambda $ ($ \hat a_{\lambda'}{}^\dagger $) is the annihilation (creation) operator and $ u_\lambda $ the appropriately normalised mode function with positive frequency.
The canonical quantisation condition to be satisfied by the operators is
\begin{equation}
  \left[\hat a^\lambda_{k_i}\,, \hat a^{\lambda'}_{k_i'}{}^\dagger\right]
  = \delta_{\lambda\lambda'}\,\delta^3(k_i-k_i')\,,
  \quad 
  \mathrm{otherwise}
  \quad
  0\,.
\end{equation}
We decide the normalisation of the modes $ u_\lambda $ by the condition for the Wronskian matrix determinant $ W $ as
\begin{equation}
  W\left[u_\lambda\,, \frac{\mathrm du_\lambda }{\mathrm d\tau}\right]
  = u_\lambda\,\frac{\mathrm du_\lambda^*}{\mathrm d\tau} 
  - u_\lambda^*\,\frac{\mathrm du_\lambda}{\mathrm d\tau} 
  = \mathrm i\,.
\end{equation}

In order to estimate the expectation value of fluctuation, we define the quantum vacuum by
\begin{equation}
  \hat a^\lambda_{k_i} |0\rangle
  = 0\,,
\end{equation}
and we construct the Fock space in the standard way.

\section{\label{sec:PGWs}Primordial gravitational waves in triaxial KdS}

In this section, we apply the formulation for quantising PGWs in triaxial KdS backgrounds.
Throughout the numerical calculations, the background anisotropy is fixed and specified by $ (q_1,q_2,q_3) = (1/\sqrt 3,0,-1/\sqrt 3) $ ($ \Theta = 8\pi/6 $), and the time of quantisation is $ t_* = 10^{-5}\,H_\Lambda^{-1} $ unless otherwise stated.
This value of $ t_* $ corresponds to the Planck scale if the energy scale of isotropic inflation is around the GUT scale.

Now, one should specify a quantum state in which the amplitude of the tensor perturbations is evaluated.
However, we do not have a reliable principle for it at this stage as we noted in section~\ref{sec:conds}, and simply take the ground state $ |0\rangle $ at $ t = t_* $ as a practical example.

The vacuum expectation value of fluctuation of tensor modes are calculated as
\begin{equation}
   \langle 0| \hat E_{ij}(x^i,\tau)\,\hat E^{ij}(x^i,\tau) |0\rangle
   = \frac{1}{a^2\,f^2}\,
   \sum_{\lambda=+,\times} \int\!\frac{\mathrm d^3k_i}{(2 \pi)^3}\,
   |u_\lambda(k_i,\tau)|^2\,,
\end{equation}
where $ u_\lambda $ is the positive frequency mode appropriately normalised at the time of quantisation, $ t = t_* $\,.
The final value of the classical expectation value at the end of inflation is calculated by evolving the mode function up to $ t \gg t_\mathrm{iso} $\,, where the modes \emph{freeze out} after their horizon exit.
We shall give analytic and numerical evaluations for the gravitational-wave power spectrum and angular distribution.

\subsection{Analytic evaluation with WKB approximation}

The mode function $ u_\lambda $ can be approximated by the (zeroth-order) WKB solution 
\begin{equation}
  u_\lambda^\mathrm{WKB}(k_i,\tau)
  =
  \frac{1}{\sqrt{2\,\Omega_\lambda(k_i,\tau)}}
  \mathrm e^{-\mathrm i\,\int^\tau\!\mathrm d\tau'\,\Omega_\lambda(k_i,\tau')}\,, 
\end{equation}
as long as the so-called WKB parameter
\begin{equation}
  Q_\lambda(k_i,\tau)
  = -\frac{1}{2\,\Omega_\lambda^2}\,\left[
    \frac{1}{\Omega_\lambda}\,
    \frac{\mathrm d^2\Omega_\lambda}{\mathrm d\tau^2} 
    - \frac{3}{2}\,\left(
      \frac{1}{\Omega_\lambda} 
      \frac{\mathrm d\Omega_\lambda}{\mathrm d\tau} 
    \right)^2
  \right]
\end{equation}
is tiny.\footnote{The function $ u_\lambda^\mathrm{WKB} $ satisfies
\begin{equation*}
  \frac{\mathrm d^2u_\lambda^\mathrm{WKB}}{\mathrm d\tau^2}
  + \left[1- Q_\lambda\right]\,\Omega_\lambda^2\,u_\lambda^\mathrm{WKB}
  = 0\,.
\end{equation*}}
It is now clear from the previous analyses that modes with sufficiently large wave number $ (k_1,k_2,k_3) $ have approximately constant $ \Omega_\lambda^2 $\,, hence $ |Q_\lambda| \ll 1 $\,.
The normalisation for $ u_\lambda^\mathrm{WKB} $ is given in terms of the Wronskian matrix determinant as
\begin{equation}
  W\left[ 
    u_\lambda^\mathrm{WKB}, 
    \frac{\mathrm du_\lambda^\mathrm{WKB}}{\mathrm d\tau}
  \right] 
  = \mathrm i\,.
\end{equation}

\begin{figure}[htbp]
  \includegraphics[scale=0.7]{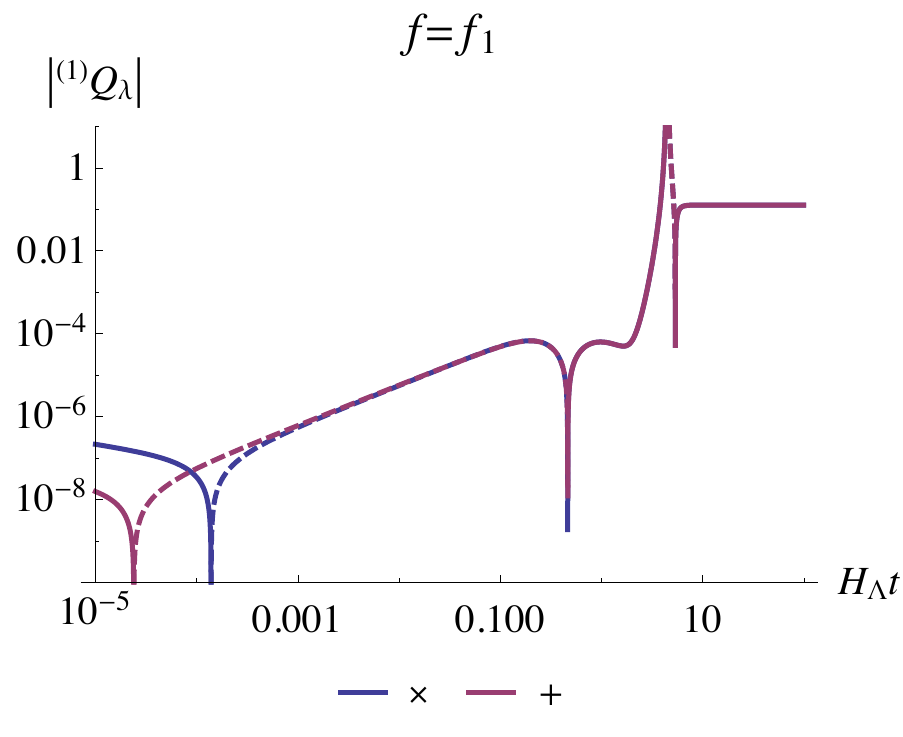}
  \includegraphics[scale=0.7]{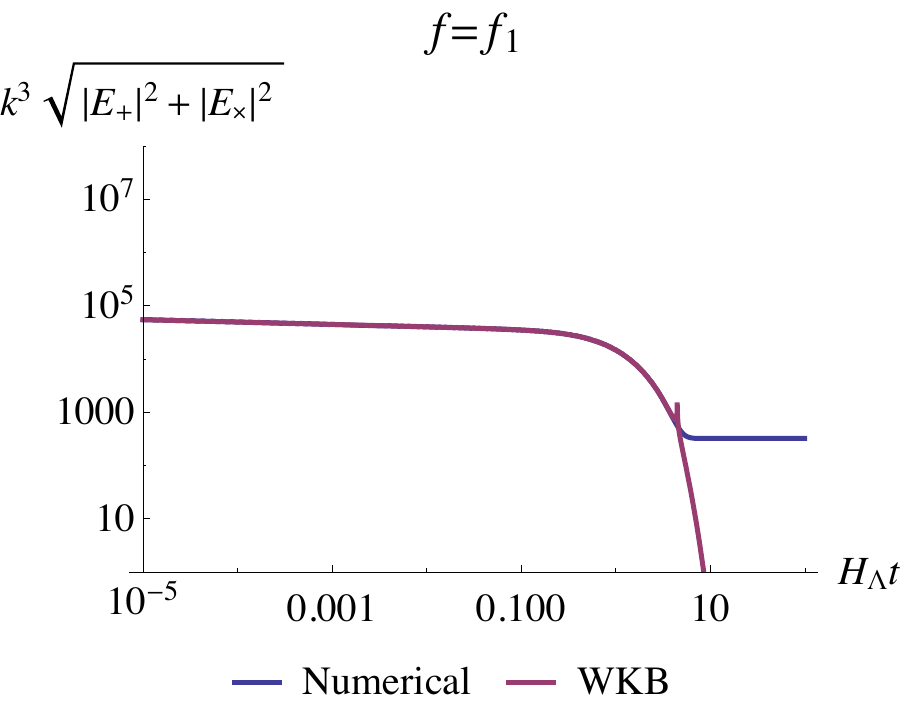}
  \caption{\label{fig:QandE}Time evolution of the WKB parameter $ {}^{(1)}Q_\lambda $ (left) and comparison of the WKB solution with the numerical solution (right) for a mode with $ k_1 = k_2 = k_3 = k/\sqrt 3 = (100/\sqrt 3)\,a_\mathrm{iso}\,H_\Lambda $\,.
    The lines for $ {}^{(1)}Q_\lambda $ are dashed when the values are negative.}
\end{figure}

As expected, for a wide range of wave number $ (k_1,k_2,k_3) $\,, the $ {}^{(1)}f $-transformed variable $ {}^{(1)}\chi_\lambda $ behaves much like a harmonic oscillator with $ {}^{(1)}\Omega_\lambda^2 \simeq k_1^2 $ as long as it does not approach too close to $ t = 0 $ or $ t = t_\mathrm{iso} $\,.
Figure~\ref{fig:QandE} shows the evolution of the WKB parameter $ |{}^{(1)}Q_\lambda| $ for $ {}^{(1)}\chi_\lambda $ and its waveform for $ k_1 = k_2 = k_3 = k/\sqrt 3 = (100/\sqrt 3)\,a_\mathrm{iso}\,H_\Lambda $\,.
The initial condition for the numerical calculation at $ t = t_* = 10^{-5}\,H_\Lambda^{-1} $ is so taken as to match the analytic WKB mode function.

As $ t $ approaches to the time of isotropisation $ t_\mathrm{iso} $\,, the WKB mode function is isotropised as
\begin{equation}
  |u_\lambda^\mathrm{WKB}|^2
  \sim
  \frac{1}{2\,\Omega_\lambda}
  \to
  \frac{1}{2\,\sqrt{k_1^2+k_2^2+k_3^2}}\,.
\end{equation}
Then, after the transition to ordinary, isotropic de Sitter inflation at $ t = t_\mathrm{iso} $\,, the tensor perturbations decay as $ |E_\lambda| \propto a^{-1} $ until they exit the horizon when $ a = k/H_\Lambda $\,.
Therefore the power spectrum of PGWs after inflation is estimated as
\begin{eqnarray}
  P_\mathrm T
  &
    \equiv
    \sum_{\lambda = +,\times} 4\,|E_{\lambda}(a=k/H_\Lambda)|^2 \nonumber\\
  &
    =
    64 \pi\,G\,\left.\frac{|u_\lambda|^2}{a^2\,f^2}\right|_{a=k/H_\Lambda} \nonumber\\
  &
    \approx
    \frac{8 \pi\,G\,H_\Lambda^2}{k^3}\,.
    \label{eq:P}
\end{eqnarray}
This expression depends only on the norm of wave number, $ k = \sqrt{k_1^2+k_2^2+k_3^2} $\,, and agrees with the one in de Sitter inflation, that is, isotropic and scale-invariant.

In the above simple estimate, we assumed that the WKB approximation is valid all the way after the time of quantisation $ t_* $ up to the time of isotropisation $ t_\mathrm{iso} $\,.
However, if the WKB condition is violated somewhere on the way during the anisotropic regime, then the prediction can differ and some imprints of the initial anisotropy could be left in the sky map of the gravitational-wave background as we studied in \cite{Furuya:2016dkh}.
Inspection of such possibilities in general backgrounds is beyond the scope of this paper and left in future studies.

\subsection{Numerical calculations: all-sky map of primordial gravitational waves}

Finally, we confirm the previous estimate for the gravitational-wave intensity by numerical means.

\begin{figure}[htbp]
  \begin{center}
    \includegraphics[scale=0.40]{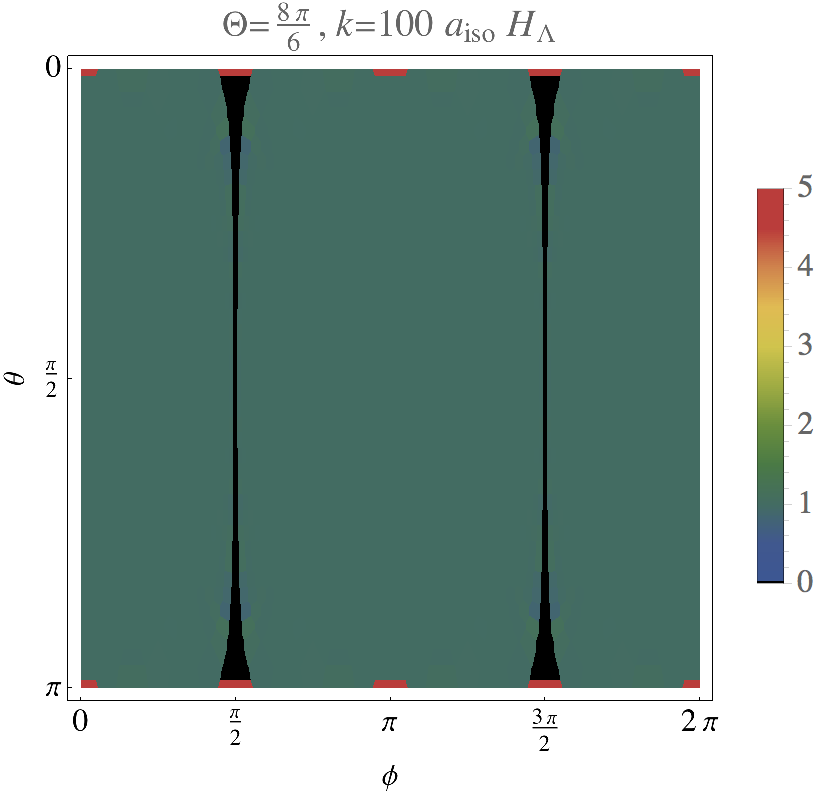}
    \includegraphics[scale=0.45]{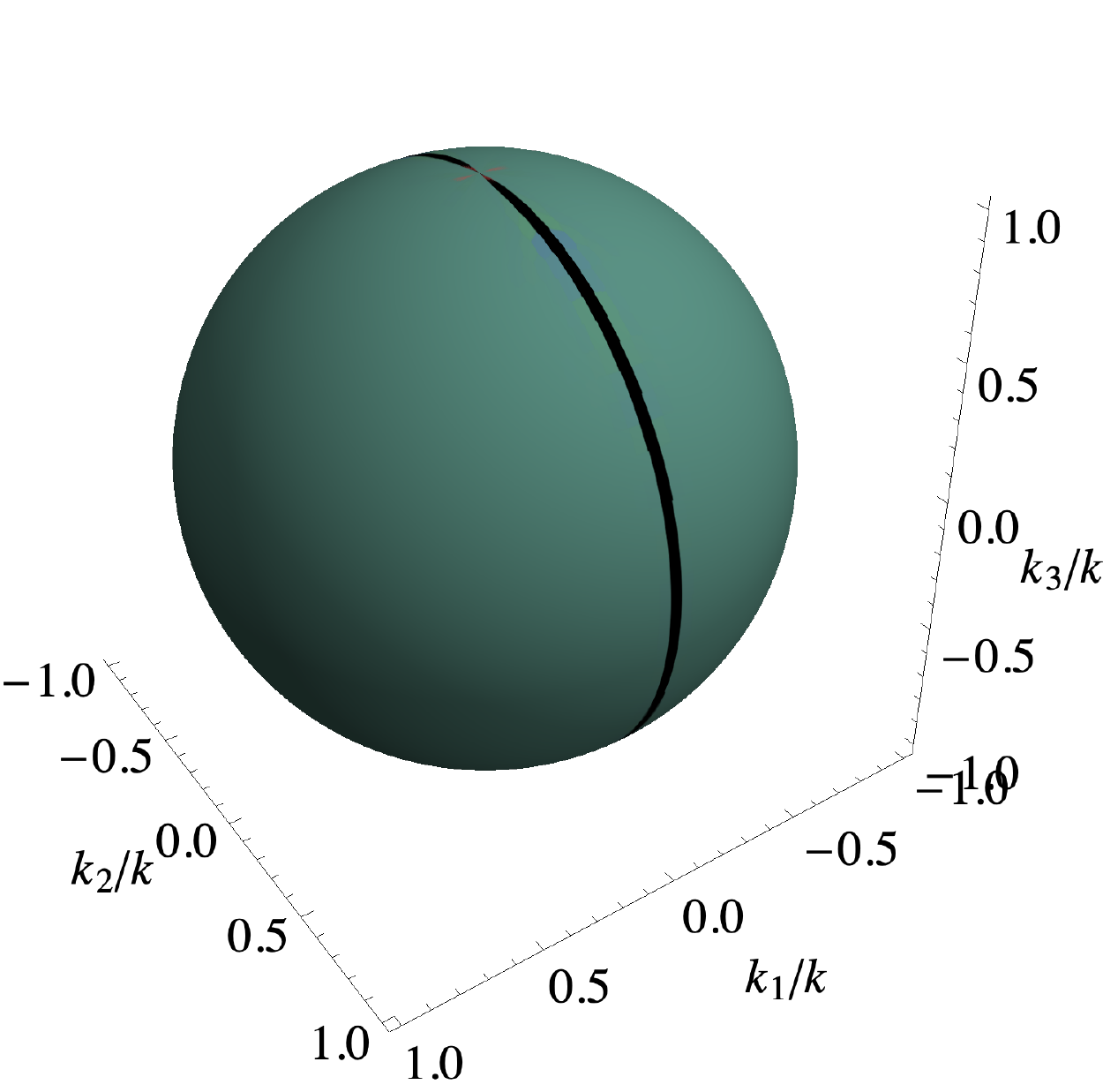} \\
    \includegraphics[scale=0.40]{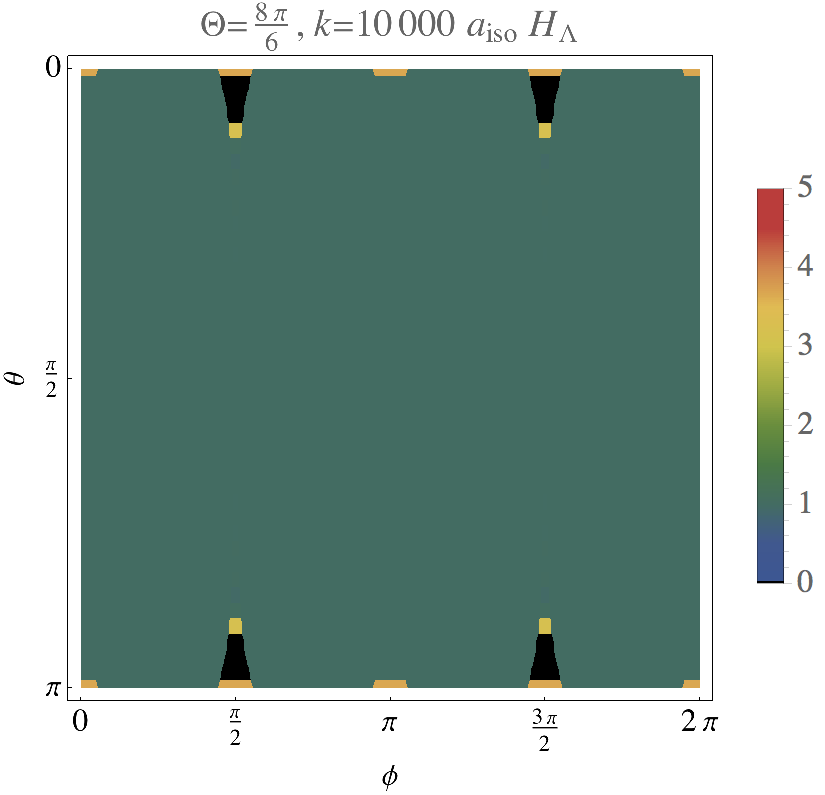}
    \includegraphics[scale=0.45]{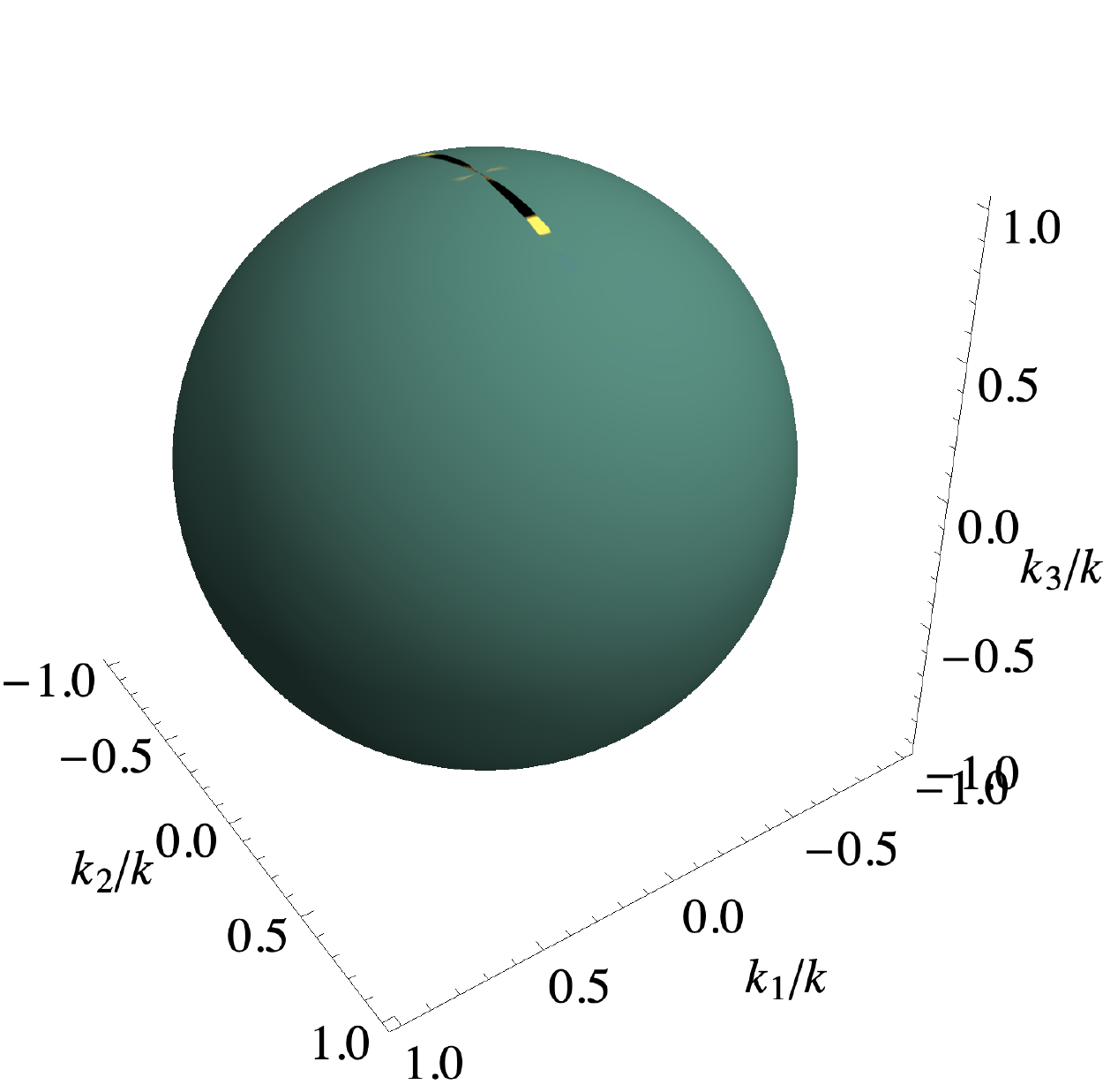}
  \end{center}
  \caption{\label{fig:map}All-sky maps of PGW intensity for wave numbers $ k = 10^2\,a_\mathrm{iso}\,H_\Lambda $ (top) and $ 10^4\,a_\mathrm{iso}\,H_\Lambda $ (bottom) in the presence of a background anisotropy characterised by the indices $ (q_1,q_2,q_3) = (1/\sqrt 3,0,-1/\sqrt 3) $ ($ \Theta = 8\pi/6 $).
    The values are normalised by those of de Sitter inflation and they are approximately unity in the vast region.
    In regions without a value (black), all $ {}^{(i)}\Omega_\times^2 $ ($ i = 1,2,3 $) are negative and we do not try to compute values.}
\end{figure}

Shown in figure~\ref{fig:map} are the all-sky maps of the intensity of PGWs for wave numbers $ k = 10^2\,a_\mathrm{iso}\,H_\Lambda $ and $ 10^4\,a_\mathrm{iso}\,H_\Lambda $\,.
The values are normalised by that of ordinary de Sitter inflation.

One can see that the prediction in the triaxial KdS coincides with that of isotropic de Sitter inflation for the range of wave length considered and wide range of direction.
This is consistent with our analytic evaluation of the power spectrum (\ref{eq:P}), and it is in principle possible to obtain the detailed power spectra of PGWs by performing similar calculations for a number of wave lengths.

There are some features, though:
In the polar regions where $ |k_3| \gg |k_1|\,, |k_2| $\,, there is an apparent enhancement of the PGWs.
At this stage, we do not claim this should be a real signature of the initial anisotropy because there vast violations of the WKB approximation occur.
The regions without value (in black) correspond to where all of $ {}^{(i)}\Omega_\lambda^2 $ ($ i=1,2,3 $) have negative values.
We do not try to compute anything in such regions since our approximation method lacks a power of prediction.
Note, however, that as we have shown in our previous paper \cite{Furuya:2016dkh}, the intensity of classical PGWs is enhanced in regions corresponding to the `uncomputable' regions in the present paper.
This may imply that there is a possibility that such regions could play an important role in detecting the imprint of the anisotropic universe in future observations.

\begin{figure}[htbp]
  \begin{center}
    \includegraphics[scale=0.40]{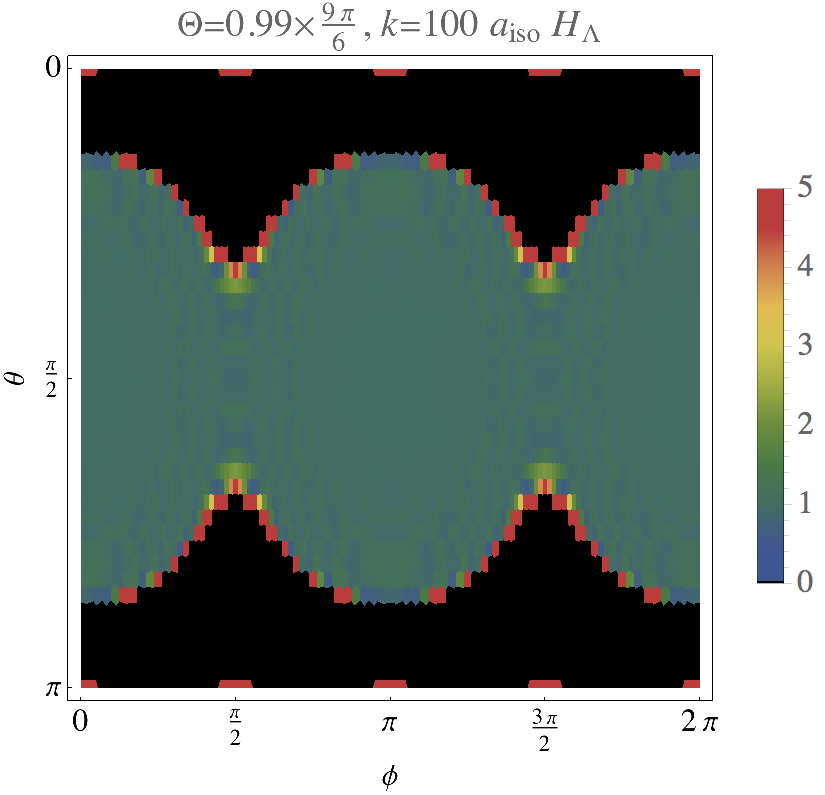}
    \includegraphics[scale=0.45]{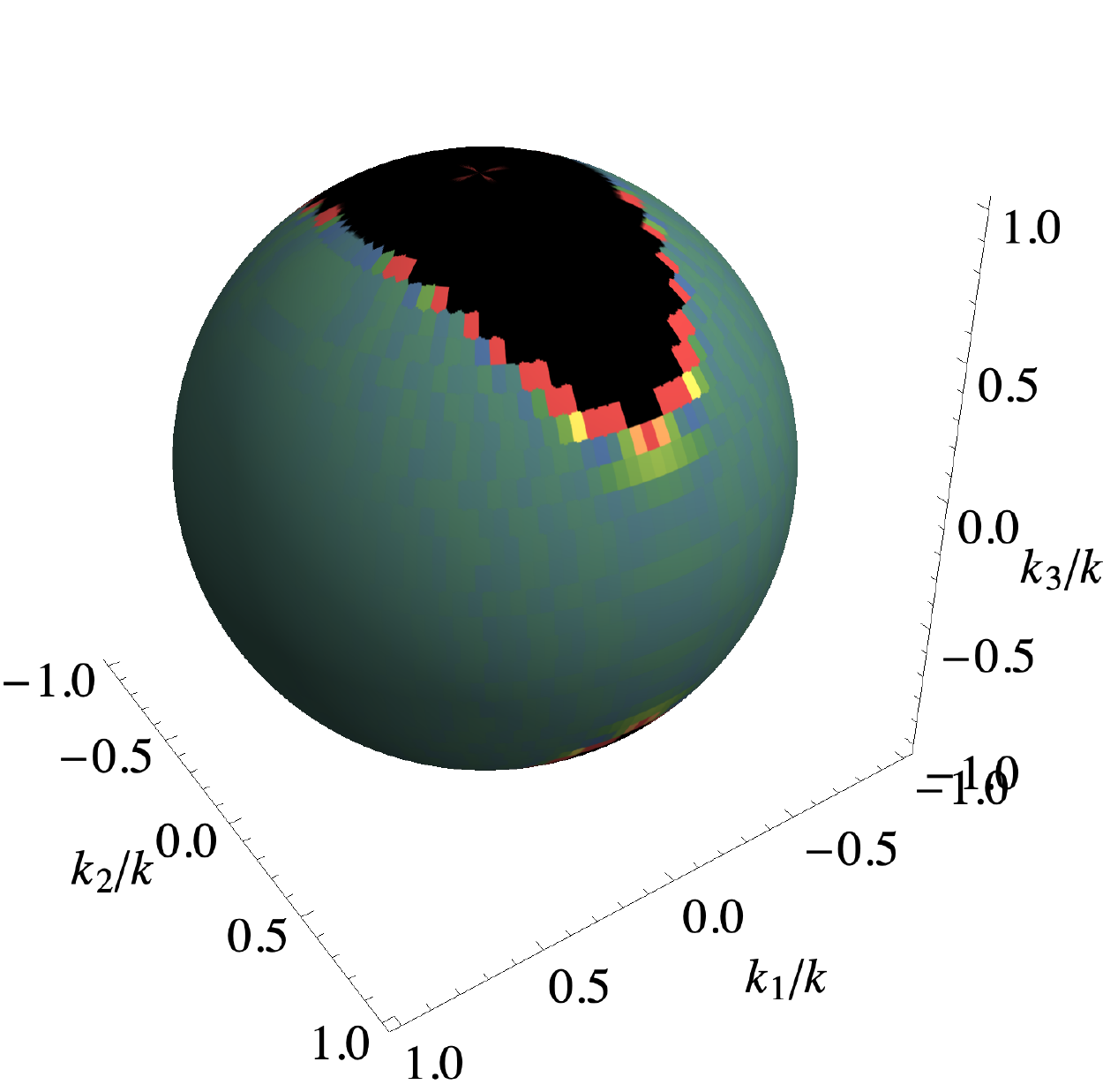}
  \end{center}
  \caption{\label{fig:mappro}All-sky map of the intensity of gravitational waves with wave number $ k = 10^2\,a_\mathrm{iso}\,H_\Lambda $ for nearly `prolate' axisymmetric background characterised by the anisotropy parameter $ \Theta = 0.99\times 9\pi/6 $\,. Compared to the top panels in figure~\ref{fig:map}, the `uncomputable' region localises near the symmetry $ k_3 $-axis. The apparent asymmetry around the $ k_3 $ axis is a consequence of the slight deviation from the exact `prolate' axisymmetry of the background.}
\end{figure}

At the end, we show a result for a different background anisotropy in order to indicate another typical result.
Figure~\ref{fig:mappro} shows the sky map for wave number $ k = 100\,a_\mathrm{iso}\,H_\Lambda $ for a nearly `prolate' axisymmetric background characterised by the anisotropy parameter $ \Theta = 0.99 \times 9\pi/6 $\,.

A conceptually subtle but reasonable difference from the triaxial case presented in the top panels of figure~\ref{fig:map} is that, in the triaxial case, the `uncomputable' region (in black) makes a great circle, while in the axisymmetric case it localises in the vicinity of the axis of symmetry, i.e., $ k_3 $-axis.
In general, for any initial anistropy, the topology of the `uncomputable' region becomes either a great-circle shape like figure~\ref{fig:map} or a pair of antipodal regions like figure~\ref{fig:mappro}.
At this stage, we do not claim it serves as an observational signature to distinguish the primordial anisotropy, because it is merely a manifestation of the limitation of our approximation method.
Still, we could direct our intention to what really happens in those regions.
We leave development of approximation method applicable in such regions in future studies.

\section{\label{sec:conc}Conclusions}

In this paper, we considered the problem of quantisation of primordial perturbations in the presence of a pre-inflationary anisotropy of Bianchi type-I.
Quantisation in such background has been considered to have a trouble due to the diverging anisotropy as approaching to the initial singularity at $ t = 0 $ except for the `oblate' axisymmetric case.

We specialised to the tensor perturbations in the Kasner--de Sitter background defined by equation~(\ref{eq:Bianchi}) with the metric functions~(\ref{eq:aandbeta}).
As the most typical example of triaxially anisotropic KdS, we took the anisotropic indices $ (q_1,q_2,q_3) = (1/\sqrt 3,0,-1/\sqrt 3) $ ($ \Theta = 8\pi/6 $) in most discussions in this paper.

In section~\ref{sec:inicon}, we began with an argument that there should be a finite initial time $ t_\mathrm{ini} $ circumventing the initial singularity.
Then we suggested that the quantum initial condition for each mode should be set at some moment $ t = t_* $ around or after $ t_\mathrm{ini} $\,, where a suitably chosen variable behaves like a harmonic oscillator and the energy spectrum of it can be determined by the standard procedure of second quantisation.
We practically examined viability of three possible variables, $ {}^{(i)}\chi_\lambda $ ($ i = 1,2,3 $), as defined in (\ref{eq:chiandtau}), for the uses as mode functions for quantisation of PGWs.
As seen in figure~\ref{fig:R}\,, due to the hierarchy $ {}^{(1)}f \gg {}^{(2)}f \gg {}^{(3)}f $\,, the variable ${}^{(1)}\chi_\lambda $ was the best to use to approximate a harmonic oscillator with a constant frequency squared $ {}^{(1)}\Omega_\lambda^2 \approx k_1^2 $ for a wide range of wave number $ (k_1,k_2,k_3) $\,.
The result also confirmed that interaction between the polarisation modes did not take significant effects at short wave lengths (see \cite{Pereira:2007yy} for a relevant analysis).

In section~\ref{sec:PGWs}, we applied our prescription to an evaluation of PGWs taking the ground state at $ t = t_* $ as the quantum state of PGWs.
For sufficiently short wave-lengths, the predicted PGWs with our quantum initial conditions gave the almost isotropic, scale-invariant power spectrum with the same amplitude as de Sitter inflation, as understood by equation~(\ref{eq:P}) and figure~\ref{fig:map}.
This is consistent with the previous results of the analyses in the short wave-length regime \cite{Pereira:2007yy,Pitrou:2008gk}.
We also presented the sky map in a nearly `prolate' axisymmetric background in figure~\ref{fig:mappro}.

For very small $ |k_1| $ or in the long wave-length regime with $ k \lesssim \mathcal O(10)\,a_\mathrm{iso}\,H_\Lambda $\,, our variables $ {}^{(i)}\chi_\lambda $ do not give a good approximation to harmonic oscillator.\footnote{Note that modes with $ k < \mathcal O(1)\,a_\mathrm{iso}\,H_\Lambda $ are already on super-horizon scales at the beginning of de Sitter inflation. Classicality of such modes might be questioned.}
The enhancement of PGWs near the $ k_3 $ axis as seen in figure~\ref{fig:map} is most probably merely a manifestation of the limitation of our approximations with the use of the variables $ {}^{(i)}\chi_\lambda $\,.
We did not even try to make any predictions if all $ {}^{(i)}\Omega_\lambda^2 $ are negative, which takes place in the vicinity of the great circle of $ k_1 = 0 $\,.

Nevertheless, based on our previous study \cite{Furuya:2016dkh}, we expect that the prediction can differ from that of de Sitter inflation if the WKB condition was violated or the frequency squared $ \Omega_\lambda^2 $ became negative during some period in the anisotropic regime.
This could indeed happen in the regions where our current approximation method breaks down.
A thorough inspection into the modes with small $ k_1 $ (corresponding to the `planar modes' studied in the `oblate' axisymmetric case \cite{Blanco-Pillado:2015dfa,Dey:2011mj,Dey:2012qp,Dey:2013tfa}) and long-wave length modes with $ k \sim \mathcal O(10)\,a_\mathrm{iso}\,H_\Lambda $ is beyond the scope of the present paper, and we leave studies of possible signatures of primordial anisotropy in the sky map of PGWs as a future work.

From the theoretical perspective, one can expect that some quantum gravity corrections would resolve the initial curvature singularity.
If it is the case, then the background geometry at the earliest stage should be modified from the Kasner space, and the results would be significantly modified.
Inspection of such possibilities is also left as a future work.

\ack
YF thanks the organisers and participants of the Workshop on Gravity and Cosmology for Young Researchers (YITP-X-16-10) held at Yukawa Institute for Theoretical Physics, Kyoto University, for stimulating discussions.
Discussions with Masato Minamitsuji were particularly useful.
He also acknowledges hospitality received from KEK during the KEK-Cosmo 2018 Workshop.
The work of YS is in part supported by JSPS KAKENHI Grant Number 16K17675.

\section*{References}

\bibliographystyle{iopart-num}
\bibliography{aniso}

\end{document}